\documentclass[useAMS,usenatbib]{mn2e}
\usepackage{graphicx}

\newcommand\araa{ARA\&A}%
\newcommand\apj{ApJ}%
\newcommand\aj{AJ}%
\newcommand\aap{A\&A}%
\newcommand\mnras{MNRAS}%
\newcommand\apjl{ApJL}%

\title[Globular cluster formation in galaxy mergers]{High-resolution simulations of galaxy mergers:\\
Resolving globular cluster formation}
\author[Bournaud et al.]{F. Bournaud$^{1,2}$\thanks{E-mail: frederic.bournaud@cea.fr}, P.-A. Duc$^{2,1}$ \& E. Emsellem$^{3}$ \\
$^{1}$CEA, IRFU, SAp, F-91191 Gif-sur-Yvette, France\\
$^{2}$Laboratoire AIM, CNRS, CEA/DSM, Universit\'e Paris Diderot. F-91191 Gif-sur-Yvette, France.\\
$^{3}$Universit\'e de Lyon, Lyon, F-69003, France ; Universit\'e Lyon~1,
Observatoire de Lyon, 9 avenue Charles Andr\'e, Saint-Genis
Laval, F-69230, France\\ CNRS, UMR 5574, Centre de Recherche Astrophysique
de Lyon ; Ecole Normale Sup\'erieure de Lyon, Lyon, F-69007, France\\
}
\begin{document}

\date{Accepted 2008 June 5.  Received 2008 May 13; in original form 2008 April 18}

\maketitle

\begin{abstract}
Massive star clusters observed in galaxy mergers are often suggested to be progenitors of globular clusters. To study this hypothesis, we performed the highest resolution simulation of a gas-rich galaxy merger so far. The formation of massive star clusters of $10^5$ to $10^7$~M$_{\sun}$, triggered by the galaxy interaction, is directly resolved in this model. We show that these clusters are tightly bound structures with little net rotation, due to evolve into compact long-lived stellar systems. Massive clusters formed in galaxy mergers are thus robust candidates for progenitors of long-lived globular clusters. The simulated cluster mass spectrum is consistent with theory and observations. Tidal dwarf galaxies of $10^{8-9}$~M$_{\sun}$ can form at the same time, and appear to be part of a different class of objects, being more extended and rotating.
\end{abstract}


\section{Introduction}

Globular clusters (GCs) are an important fossil record of the evolution of physical conditions in galaxies \citep{bekki08}, because the formation of massive clusters is triggered by shocks and high pressures in the interstellar medium \citep{vdb79,AZ01}. Fundamental properties along the Hubble sequence are a higher frequency of GCs around elliptical than disk galaxies \citep{harris91} and a bimodal population in particular around early-type galaxies, with a population of low-metallicity GCs and a population of younger, higher-metallicity ones \citep{AZ92}.

Since the works of \citet{schweizer} and \citet{whitmore}, there has been increasing evidence that young massive star clusters (YMC) form in interacting and merging galaxies, and these are often proposed to be GC progenitors. Theoretically, massive clusters form in mergers because of shocks and high turbulent pressure in interacting galaxies, which favors the formation of tightly bound clusters rather than unbound associations \citep{elmegreen-efremov1997}. If GCs can form this way, given that most elliptical galaxies formed by galaxy mergers (Naab \& Burkert 2003, Bournaud et al. 2005), this mechanism could account for the youngest and most metallic GCs that are more frequent around early-type galaxies \citep{AZ01}. These GCs would come in addition to those formed early in the Universe as a result of thermal instabilities in proto-galaxies \citep{fall-rees85}, strong shocks at the epoch of the Reionization \citep{cen01}, or stripping of nucleated dwarf galaxies \citep{freeman90,gao}. 
Nevertheless, that an important population of GCs formed in galaxy mergers is still challenged by some observations \citep{spitler}.
Whether or not YMCs formed during mergers contribute to the present-day GC populations is still an open question also because the long-term evolution of such YMCs is uncertain \citep[see][]{degrijs07}. A requirement for YMCs to become long-lived GCs is that they should be gravitationally bound, which is difficult to assess observationally, as the velocity dispersion of a cluster does not directly trace its mass because of mass segregation \citep{fleck06}. The survival issue is further complicated by the the mass-loss of clusters if their IMF is too shallow, and by the tidal field of the parent galaxy which may progressively disrupt these clusters \citep{miocchi}. 

Numerical simulations are a powerful tool to study galaxy mergers, but resolving the formation of star clusters in self-consistent models requires a huge dynamical range, because structures smaller than 100~pc should be resolved in a simulated volume larger than 100~kpc. Models by \citet{bekki02} and \citet{KG} have shown that the pressure and density required to form massive bound star clusters could be reached in galaxy mergers. They could however not identify individually forming GCs. \citet{limaclow} studied individual GC formation in an indirect manner, using a model where absorbing sink particles are assumed to form above a chosen density threshold. This allows the mass and spatial distribution of the putative GCs to be studied. However, the absorbing nature of sink particles in such a model dooms any stellar association to be endlessly bound, so that whether or not long-lived bound objects can actually form in mergers is not probed by such a model.

In this Letter, we present a simulation of a wet galaxy merger, which to our knowledge is the highest resolution model of this kind so far. We directly resolve the formation of dense structures resembling Super Star Clusters (SSCs) with typical masses of $10^{5-7}$~M$_{\sun}$. Comparing their gravitational and kinetic energy, we argue that they are tightly bound and likely progenitors of long-lived GCs. 

\section{Numerical simulations}

We use a particle-mesh code to compute the gravitational dynamics of gas, stars, and dark matter \citep{bournaud05}. The grid cell size is 32~pc up to 25~kpc from each galaxy center, 64~pc up to radii of 50~kpc, and 128~pc at larger radii. The density is computed with a Cloud-in-Cell interpolation, and a FFT technique is used to compute the gravitational potential, with a Plummer softening length of 32~pc for all types of particles. Gas dynamics is modeled with a sticky-particle scheme with elasticity parameters $\beta _t$=$\beta_r$=0.6. The star formation rate is computed using a Schmidt-Kennicutt law: the star formation rate is proportional to the gas density in each cell to the exponent 1.5. Gas particles are converted to star particles with a corresponding rate in each cell. Energy feedback from supernovae is accounted for with the scheme proposed by \citet{MH94}. Each stellar particle formed has a number of supernovae computed from the fraction of stars above 8~M$_{\sun}$ in a Miller-Scalo IMF. A fraction $\epsilon$ of the $10^{51}$~erg energy of each supernova is released in the form of radial velocity kicks applied to gas particles within the closest cells. We use $\epsilon = 2 \times 10^{-4}$, as \citet{MH94} suggest that realistic values lie around $10^{-4}$ and less than $10^{-3}$.

We use a total number of particles of 36 millions, corresponding to 6 million particles per galaxy per component (gas, stars, and dark matter). The particle mass is $3.4 \times 10^4$~M$_{\sun}$ for initially present stars and $7\times 10 ^3$~M$_{\sun}$ for gas and stars formed during the simulation. To our knowledge, this is the highest-resolution simulation of a wet galaxy merger published so far. \citet{wetzstein} performed a merger simulation with $4\times 10^6$ particles per galaxy, but only 45,000 particles for the gas component. The simulation of early-type galaxy formation by \citet{naab} uses $8 \times 10^6$ particles, but the large volume modeled in this cosmological run limits the resolution (softening) to 250~pc. \citet{limaclow} had $5 \times 10^5$ gas particles per galaxy, which is one order of magnitude below our number and gives an equivalent particle mass only if one assumes a much lower galactic mass; they use a softening of 10~pc for gas particles but 100~pc for stellar particles, so that gravity is only modeled accurately at scales larger than 100~pc since the gravity of gas and stars cannot be decoupled \citep[e.g.,][]{jog}.

The initial setup was chosen to be representative of an equal-mass wet merger at low or moderate redshift (0$\leq$z$\leq$1) with a gas fraction of 17\%. Each spiral galaxy is made-up of a stellar disk with a Toomre profile of scale-length 4~kpc truncated at 10~kpc, and a gas disk of scale-length 8~kpc truncated at 20~kpc, which models the extended HI disks detected well beyond the optical edge in most spirals \citep{RH94}. The bulge has a Plummer profile of scale-length 0.8~kpc. The dark halo has a Burkert profile with a 8~kpc core radius, truncated at 75~kpc. The initial stellar mass is $2 \times 10^{11}$~M$_{\sun}$, 18\% of this mass being in the bulge. The gas disk mass is $4 \times 10^{10}$~M$_{\sun}$. The dark halo mass is $5.5\times 10^{11}$~M$_{\sun}$.
The orbit in the merger model was chosen to avoid peculiar situations like coplanar disks or head-on encounters. It is prograde for one galaxy, with an angle between the disk and orbital plane of 30 degrees. It is retrograde for the other galaxy, and the angle between its disk and the orbital plane of 70 degrees. The pericenter distance is 25~kpc and the velocity at infinite distance is 150~km~s$^{-1}$.  

\section{Results}


\begin{figure*}
\centering
\includegraphics[width=16.5cm]{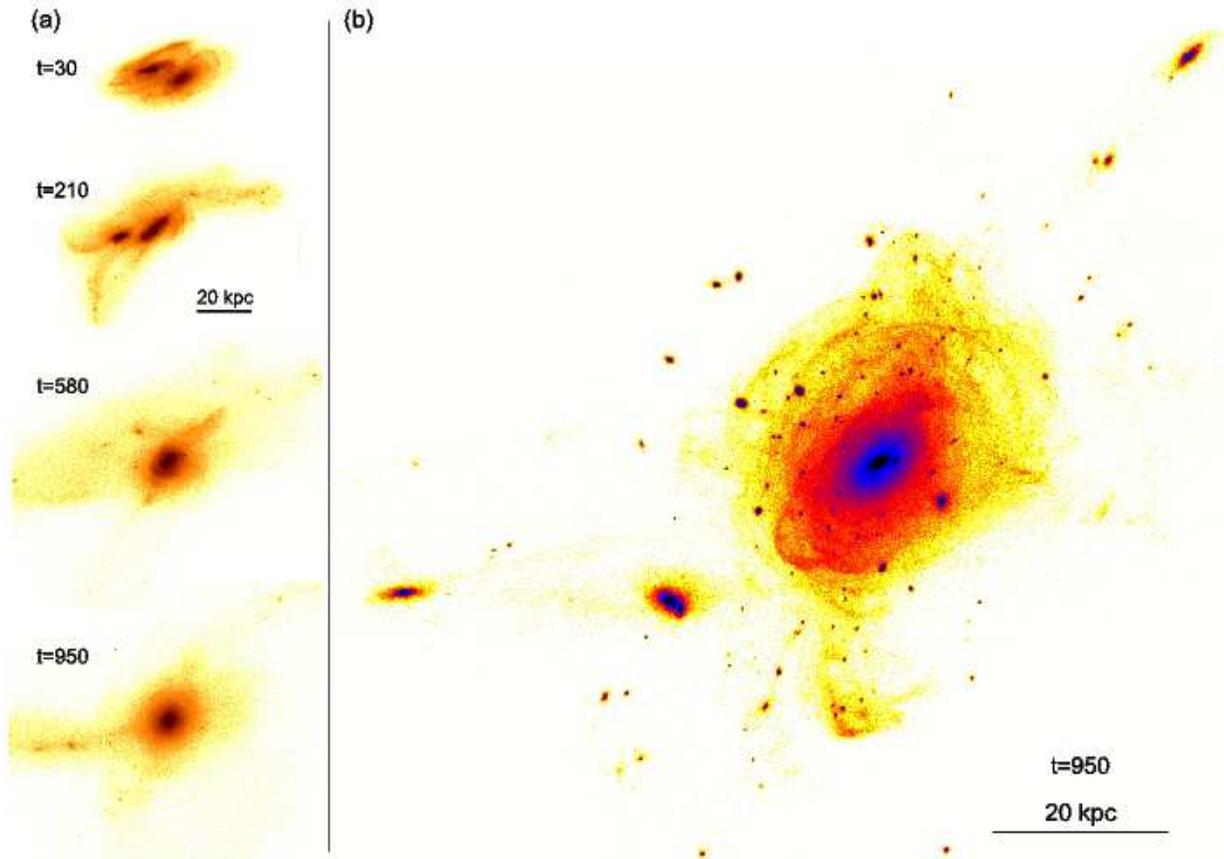}
\caption{{\bf (a)} Sequence of four snapshots showing the merger evolution. Time is indicated in Myr after the first pericenter passage. The total stellar mass density is shown -- {\bf (b)} Surface density of ``young'' stars (defined as those formed after the first pericenter passage) at the end of the wet merger simulation, 950~Myr after the first pericenter passage. Compact SSCs are seen all around the central elliptical galaxy, together with three massive tidal dwarfs in the remnants of the two long tidal tails.}\label{fig:ystar} \label{fig:sequence}
\end{figure*}

\subsection{SSCs in isolated and merger models}

The evolution of the merging galaxy pair is shown on Fig.~\ref{fig:sequence}. Both galaxies develop tidal tails shortly after the pericenter passage while their central bodies merge together in less than 500~Myrs. One billion year after the first pericenter passage of the two spiral progenitors, the central elliptical galaxy is already relaxed, while tidal debris are still visible around it. As in all mergers of this type, the remnant has the shape, profile and kinematics of an elliptical galaxy.
At the end of the simulation, the morphology resembles that of observed late-stage mergers, like IC~1182 on the last snapshot of panel (a) or NGC~7252 on panel (b) in Fig.~\ref{fig:ystar}.

Shells and filaments are present as expected for a young merger remnant. Several compact and dense structures are visible in and around the elliptical remnant, over a large radial range. The most massive are manifest in the total stellar mass distribution, and the others are emphasized by plotting the density of the ``young'' stars alone (defined as those formed after the first pericenter passage). More than one hundred objects of this type are visible, with stellar masses in the $10^{5-7}$~M$_{\sun}$ range, diameters between 10 and 100~pc. They are made-up almost exclusively of ``young'' stars and as shown below are gravitationally bound: they should therefore be considered as SSCs formed during the galaxy merger. We will show later that the three most massive objects, above $10^8$~M$_{\sun}$, should be considered as systems of another class, namely Tidal Dwarf Galaxies.

\begin{figure}
\centering
\includegraphics[width=7cm]{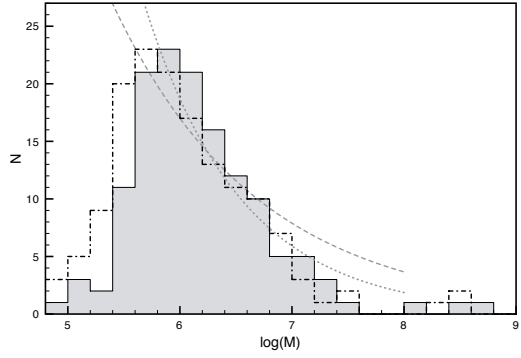}
\caption{Mass spectrum of the stellar structures formed in the merger model. The dashed histogram is for the test simulation with reversed sticky particle parameters. Power-law spectra with slopes -2 (dashed) and -3 (dotted) are shown.}\label{fig:mspec}
\end{figure}

\begin{figure}
\centering
\includegraphics[width=5.2cm]{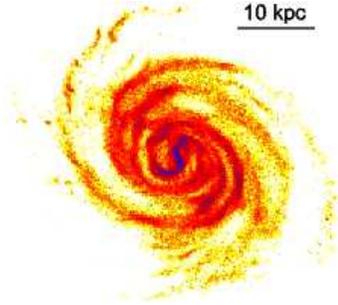}
\caption{Isolated galaxy evolved over the same duration as the merger. The young stellar density is shown and does not harbor substructures as dense and compact as in the merger remnant.}\label
{fig:control}
\end{figure}

\begin{figure}
\centering
\includegraphics[width=7cm]{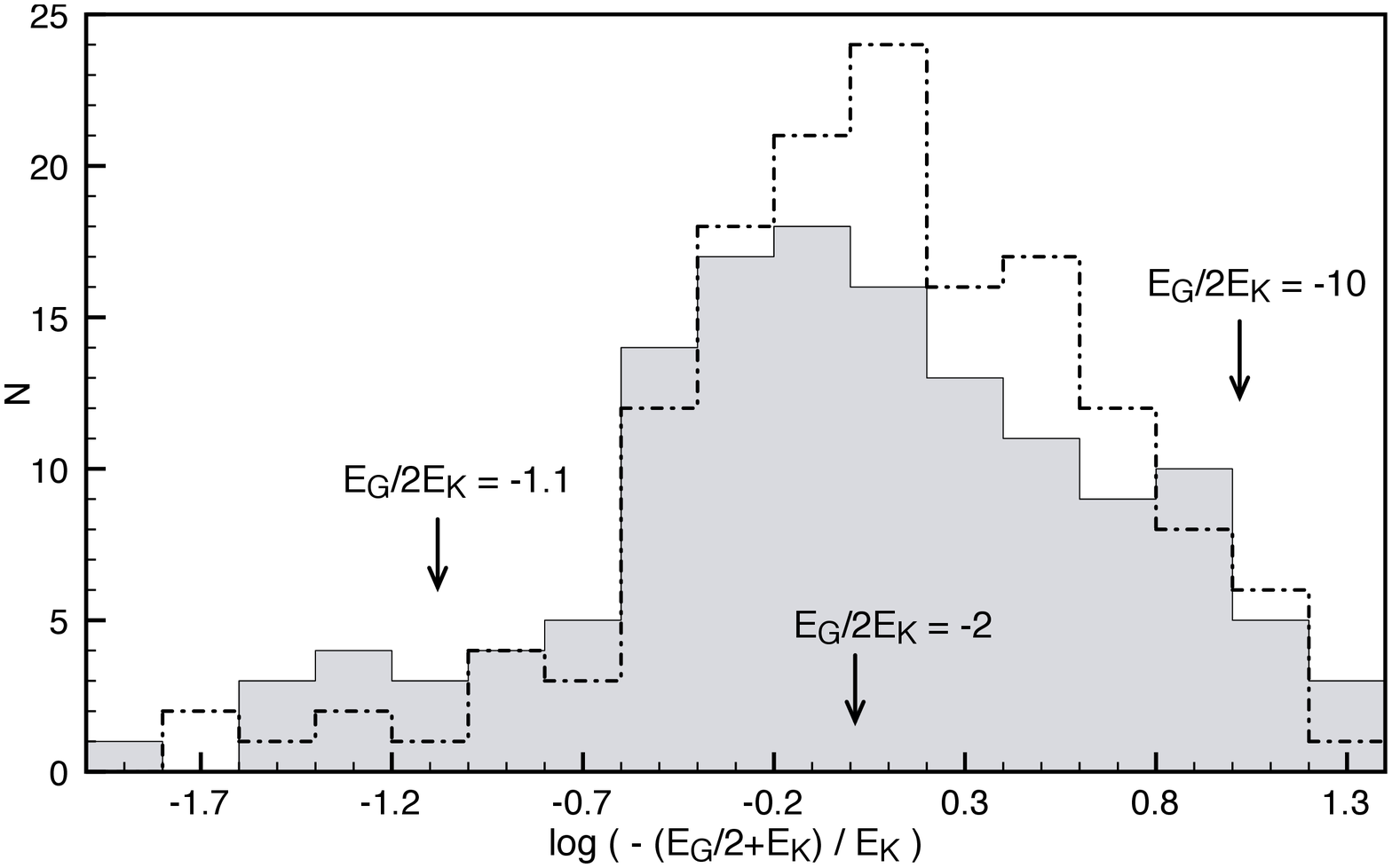}
\caption{Distribution of $1/2 E_G + E_K$ for the SSCs in our model, where $E_G$ is computed for a true, non-softened Gravity. This value would be 0 ($- \infty$ in $\log$scale) for virialized systems. Our SSCs are characterized by $E_G/2E_K <-1$, which indicates that their collapse was stopped by the spatial resolution limit; they would become even more compact at infinite resolution.}\label{fig:energy}
\end{figure}

These SSCs are, for a large part, well above the mass resolution of our models: the number of particles describing a cluster of $10^6$~M$_{\sun}$ is 140; the mass spectrum of the clusters formed during the merger is shown on Fig.~\ref{fig:mspec}. Thus, except for those of $\sim 10^5$~M$_{\sun}$, these SSCs are not remnant objects that were prevented to fragment because of a too small number of particles. To prove that the formation of these SSCs is really triggered by the galaxy merger, and not simply by a general instability in our galaxy models, we show on Fig.~\ref{fig:control} the isolated disk evolution at the same resolution as our merger model. It has been evolved for the same duration as the merger and the figure shows the ``young'' stars, defined as the stars formed after $t=300$~Myr (the pericenter in the merger model). This model does not show many dense and compact substructures. A few associations of young stars are visible in the outer spiral arms, but none reaches a mass of $10^6$~M$_{\sun}$ and they are not as compact as the structures formed during the merger; these are only short-lived associations of young stars. This shows that our isolated disk galaxy model is overall stable against the formation of SSCs. This is consistent with observations of spiral galaxies where young SSCs form mostly in the inner resonant rings of bars \citep{galliano}, while no such ring is formed in our isolated galaxy model over the Gyr-long evolution period. The formation of dense, SSC-like objects in our merger model is a direct consequence of the galaxy interaction itself. To check that the results are not crucially dependent on the sticky particle parameters, we ran another simulation with $\beta _t$=$\beta_r$=$-0.6$ instead of +0.6: colliding clouds pass through each other instead of bouncing back, which is a more dissipative case. The final system has similar SSC properties (Figs.~\ref{fig:mspec} and \ref{fig:energy}).

A noticeable property of the interstellar gas during the galaxy interaction is its increased velocity dispersions. At $t$=450, which corresponds to the star formation peak, the average dispersion of gas clouds is 16~km~s$^{-1}$ and 17\% of gas clouds experience local dispersions larger than 30~km~s$^{-1}$. In the isolated disk at same instant, the average value is 11~km~s$^{-1}$, and only 3\% of the clouds are above 30~km~s$^{-1}$. This is consistent with the large dispersions observed in interacting systems \citep{elmegreen95}. As a result, the Jeans mass is increased and more massive bound structures can form in the gas. Nevertheless, such massive structures could become unbound after star formation has operated \citep{GB01}. Actually, the gas pressure 
is enhanced during the galaxy merger, with local spikes of high pressure in the caustics and shocks caused by the interaction: the average pressure is increased by 55\% and the 95th percentile of the pressure distribution by a factor 9. Star formation in such high-pressure regions has a high efficiency, so that the resulting stellar systems can remain bound: this enables the formation of massive bound star clusters.

\subsection{Evolution into globular clusters}

\begin{figure}
\centering
\includegraphics[width=8cm]{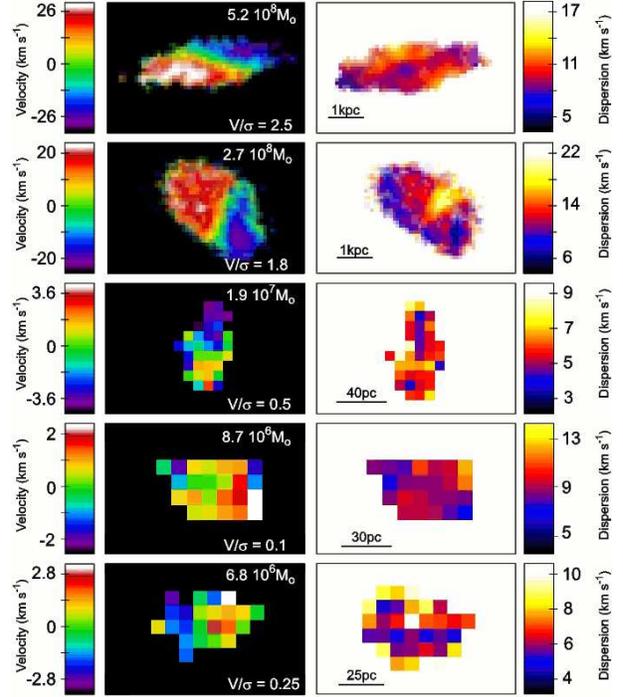}
\caption{Stellar velocity and dispersion fields for 2 TDGs and 3 SSCs. The stellar mass and typical rotation velocity to dispersion ratio ($V/\sigma$) are indicated. TDGs are rotating disks with $V/\sigma$ larger than 1. GC progenitors are dominated by random motions.}\label{fig:kin}
\end{figure}

The young SSCs observed in galaxy mergers and reproduced in our model are often suggested to be progenitors of GCs. The infant mortality can affect these SSCs but a large fraction can still be long-lived \citep{degrijs07} and may thus become GCs. This however requires that they are dense enough and tightly bound by their own gravity, which our model allows to check directly. Measuring the half-mass radius of the SSCs in our model, we find that more than two thirds have radii between 10 and 35~pc: even if they are above the mass resolution limit, they are around the spatial resolution, which suggests that they have reached an equilibrium because of the softening of Gravity at short distances. We have measured for each cluster the kinetic energy $E_K$ and gravitational energy $E_G$ for the real, non-softened gravity. While the value of $1/2 E_G + E_K$ should be zero for structures in pure gravitational equilibrium, we find that this quantity is significantly negative (equivalently $E_G/2E_K <-1$). This confirms that the clusters have reached a pseudo-equilibrium resulting from the softening, and would become more compact at infinite resolution, with half-mass radii then significantly smaller than the 10--35~pc found here.

The SSCs formed during the galaxy merger are thus robust candidates for GC progenitors. They have {\it masses} typical of GCs down to our mass limit. They are {\it compact} objects with typical half-mass radii that would be smaller than 10~pc at infinite resolution. They are {\it bound} and {\it long-lived}: many of them are already formed 500~Myr before the final snapshot and their boundedness would increase at higher resolution. 

\subsection{Mass spectrum and tidal dwarf galaxies}

The mass spectrum of young stellar objects 
is shown on Fig.~\ref{fig:mspec}. 
Masses of $10^5$~M$_{\sun}$ can be affected by the mass resolution of our models, and the turnoff of the mass function below $5\times 10^5$~M$_{\sun}$ cannot be considered as robust. At higher masses, the mass spectrum decreases with a power-law shape. 61 objects have masses of $10^{5-6}$, 66 of $10^{6-7}$, only 6 of $10^{7-8}$. There is a noticeable gap 
between the most massive SSCs and the 3 objects with masses of $10^{8-9}$~M$_{\sun}$, which resemble Tidal Dwarf Galaxies (TDGs, \citealt{bournaudduc06}). TDGs are much less concentrated than the SSCs: SSCs of a few $10^6$~M$_{\sun}$ have half-mass radii around 20~pc, the nine most massive SSCs of 1 to $5 \times 10^7$~M$_{\sun}$ have an average radius of 48~pc, while the three TDGs have radii of 1-2~kpc. This leads to typical surface densities ($\propto M/R^2$) one order of magnitude lower for the TDGs. While the SSCs/GC-progenitors are compact structures supported by random motions with little net rotation and $V$/$\sigma$ ratios much smaller than 1, TDGs have a large angular momentum ($V$/$\sigma$$\simeq$2-3) and take the form of rotating gas disks (see Fig.~\ref{fig:kin}) as also found in observations (e.g. \citealt{bournaud07}). 

Because of these physical differences, in addition to the possible gap in the mass function, our model suggests that two families of objects can form in galaxy mergers. Compact SSCs, in which rapid star formation occurs, are robust candidates for GC progenitors. TDGs are not simply the high-mass end of the SSC distribution: they are made-up of rotating disks forming stars at a more moderate pace, and will likely evolve into dwarf satellite galaxies \citep{bournaudduc06, metzkroupa07}. The SSCs/GCs form locally by the gravitational collapse of overdense regions in the turbulent interstellar medium, following the mechanism studied for instance by \citet{wetzstein}, while TDGs are formed in regions where large amounts have been accumulated during the interaction or in the progenitor disks, as proposed by \citet{elmegreen93} and \citet{duc04}.

As for the mass spectrum of SSCs themselves, above the mass limit and up to a few $10^7$~M$_{\sun}$, the best-fitting slope in a $N\propto M^{\alpha}$ model would be $\alpha \simeq -3$, as shown on Fig.~\ref{fig:mspec}. However, the -2 slope expected from theoretical models \citep{elmegreen-efremov1997} is compatible with the bulk of the mass distribution, assuming that the spectrum is truncated around $10^7$~M$_{\sun}$, which may indeed be realistic and could result from various physical effects \citep{gieles}. Models spanning a larger mass range or repeated runs to increase the statistics are be required to firmly assess the shape of the cluster mass function. Nevertheless, our present model is compatible with observations suggesting a slope around -2 or steeper \citep{weidner}.

\section{Summary}

Using a high-resolution simulation of a galaxy merger, we resolve the formation of structures down to masses of $10^5$~M${\sun}$. This enables us to directly reproduce the formation of numerous Super Star Clusters, as observed in merging galaxies. We find that these are dense, tightly bound structures that will evolve into compact stellar systems and are likely progenitors of numerous globular clusters (GC). Tidal dwarf galaxies of a few $10^8$~M$_{\sun}$ are also formed in our model; they appear to be a different type of objects, less concentrated and resolved as rotating disks. Our results suggest that both TDGs and SSCs can form at the same time in galaxy mergers. The mass function of GC progenitors in this model has a power-law shape with a slope between -2 and -3, likely truncated around $10^7$~M$_{\sun}$. The formation of globular cluster is much more efficient in merging galaxies than in isolated disk models. The total mass of the GC progenitors in our merger model amounts to 4\% of the available gas mass, which is 0.7\% of the total baryonic mass. The efficiency could be even higher with the large gas fractions observed in high-redshift disks \citep{daddi}. GC progenitors form in the tails and caustics with high turbulent speed and pressure around the central merger remnant. The material in such regions has been expulsed from the spiral disks where it has been previously enriched; it is known to have relatively high metallicities \citep{weilbacher03}, so the GCs formed this way should also be metal-rich. The excess of GCs around elliptical and lenticular galaxies compared to spirals could then be explained by this mechanism, given that mergers can form early-type galaxies and GCs at the same time.

\section*{Acknowledgments}
Enlightening discussions with Bruce Elmegreen, Glenn van de Ven, Richard de~Grijs and Pavel Kroupa are greatly appreciated, as well as useful comments by the referee. Simulations were carried out on the NEC-SX8R at CEA/CCRT. 



\begin{thebibliography}{38}
\expandafter\ifx\csname natexlab\endcsname\relax\def\natexlab#1{#1}\fi

\bibitem[{{Ashman} \& {Zepf}(1992)}]{AZ92}
{Ashman} K.~M., {Zepf} S.~E., 1992, \apj, 384, 50

\bibitem[{{Ashman} \& {Zepf}(2001)}]{AZ01}
---, 2001, \aj, 122, 1888

\bibitem[{{Bekki} {et~al.}(2002){Bekki}, {Forbes}, {Beasley}, \&
  {Couch}}]{bekki02}
{Bekki} K., {Forbes} D.~A., {Beasley} M.~A., {Couch} W.~J., 2002, \mnras, 335,
  1176

\bibitem[{{Bekki} {et~al.}(2008){Bekki}, {Yahagi}, {Nagashima}, \&
  {Forbes}}]{bekki08}
{Bekki} K., {Yahagi} H., {Nagashima} M., {Forbes} D.~A., 2008,
  astro-ph:0804.1842

\bibitem[{{Bournaud} \& {Duc}(2006)}]{bournaudduc06}
{Bournaud} F., {Duc} P.-A., 2006, \aap, 456, 481

\bibitem[{{Bournaud} {et~al.}(2007){Bournaud}, {Duc}, {Brinks}, {Boquien},
  {Amram}, {Lisenfeld}, {Koribalski}, {Walter}, \& {Charmandaris}}]{bournaud07}
{Bournaud} F., {et~al.} 2007, Science, 316,
  1166

\bibitem[{{Bournaud} {et~al.}(2005){Bournaud}, {Jog}, \& {Combes}}]{bournaud05}
{Bournaud} F., {Jog} C.~J., {Combes} F., 2005, \aap, 437, 69

\bibitem[{{Cen}(2001)}]{cen01}
{Cen} R., 2001, \apj, 560, 592

\bibitem[{{Daddi} {et~al.}(2008){Daddi}, {Dannerbauer}, {Elbaz}, {Dickinson},
  {Morrison}, {Stern}, \& {Ravindranath}}]{daddi}
{Daddi} E., {Dannerbauer} H., {Elbaz} D., {Dickinson} M., {Morrison} G.,
  {Stern} D., {Ravindranath} S., 2008, \apjl, 673, L21

\bibitem[{{de Grijs}(2007)}]{degrijs07}
{de Grijs} R., 2007, in Young massive star clusters - Initial conditions and
  environments, E. Perez, R. de Grijs, R. M. Gonzalez Delgado, eds.
  astro-ph:0711.3540

\bibitem[{{Duc} {et~al.}(2004){Duc}, {Bournaud}, \& {Masset}}]{duc04}
{Duc} P.-A., {Bournaud} F., {Masset} F., 2004, \aap, 427, 803

\bibitem[{{Elmegreen} \& {Efremov}(1997)}]{elmegreen-efremov1997}
{Elmegreen} B.~G., {Efremov} Y.~N., 1997, \apj, 480, 235

\bibitem[{{Elmegreen} {et~al.}(1993){Elmegreen}, {Kaufman}, \&
  {Thomasson}}]{elmegreen93}
{Elmegreen} B.~G., {Kaufman} M., {Thomasson} M., 1993, \apj, 412, 90

\bibitem[{{Elmegreen} {et~al.}(1995){Elmegreen}, {Kaufman}, {Brinks},
  {Elmegreen}, \& {Sundin}}]{elmegreen95}
{Elmegreen} D.~M., {Kaufman} M., {Brinks} E., {Elmegreen} B.~G., {Sundin} M.,
  1995, \apj, 453, 100

\bibitem[{{Fall} \& {Rees}(1985)}]{fall-rees85}
{Fall} S.~M., {Rees} M.~J., 1985, \apj, 298, 18

\bibitem[{{Fleck} {et~al.}(2006){Fleck}, {Boily}, {Lan{\c c}on}, \&
  {Deiters}}]{fleck06}
{Fleck} J.-J., {Boily} C.~M., {Lan{\c c}on} A., {Deiters} S., 2006, \mnras,
  369, 1392

\bibitem[{{Freeman}(1990)}]{freeman90}
{Freeman} K.~C., 1990, {Our fossil Galaxy.}, Dynamics and Interactions of
  Galaxies, pp. 36--47

\bibitem[{{Galliano} {et~al.}(2005){Galliano}, {Alloin}, {Pantin}, {Lagage}, \&
  {Marco}}]{galliano}
{Galliano} E., {Alloin} D., {Pantin} E., {Lagage} P.~O., {Marco} O., 2005,
  \aap, 438, 803

\bibitem[{{Gao} {et~al.}(2007){Gao}, {Jiang}, \& {Zhao}}]{gao}
{Gao} S., {Jiang} B.-W., {Zhao} Y.-H., 2007, ChJAA, 7, 111

\bibitem[{{Geyer} \& {Burkert}(2001)}]{GB01}
{Geyer} M.~P., {Burkert} A., 2001, \mnras, 323, 988

\bibitem[{{Gieles} {et~al.}(2006){Gieles}, {Larsen}, {Scheepmaker}, {Bastian},
  {Haas}, \& {Lamers}}]{gieles}
{Gieles} M., {Larsen} S.~S., {Scheepmaker} R.~A., {Bastian} N., {Haas} M.~R.,
  {Lamers} H.~J.~G.~L.~M., 2006, \aap, 446, L9

\bibitem[{{Harris}(1991)}]{harris91}
{Harris} W.~E., 1991, \araa, 29, 543

\bibitem[{{Jog}(1996)}]{jog}
{Jog} C.~J., 1996, \mnras, 278, 209

\bibitem[{{Kravtsov} \& {Gnedin}(2005)}]{KG}
{Kravtsov} A.~V., {Gnedin} O.~Y., 2005, \apj, 623, 650

\bibitem[{{Li} {et~al.}(2004){Li}, {Mac Low}, \& {Klessen}}]{limaclow}
{Li} Y., {Mac Low} M.-M., {Klessen} R.~S., 2004, \apjl, 614, L29

\bibitem[{{Metz} \& {Kroupa}(2007)}]{metzkroupa07}
{Metz} M., {Kroupa} P., 2007, \mnras, 376, 387

\bibitem[{{Mihos} \& {Hernquist}(1994)}]{MH94}
{Mihos} J.~C., {Hernquist} L., 1994, \apj, 437, 611

\bibitem[{{Miocchi} {et~al.}(2006){Miocchi}, {Capuzzo Dolcetta}, {Di Matteo},
  \& {Vicari}}]{miocchi}
{Miocchi} P., {Capuzzo Dolcetta} R., {Di Matteo} P., {Vicari} A., 2006, \apj,
  644, 940

\bibitem[{{Naab} \& {Burkert}(2003)}]{NB03}
{Naab} T., {Burkert} A., 2003, \apj, 597, 893

\bibitem[{{Naab} {et~al.}(2007){Naab}, {Johansson}, {Ostriker}, \&
  {Efstathiou}}]{naab}
{Naab} T., {Johansson} P.~H., {Ostriker} J.~P., {Efstathiou} G., 2007, \apj,
  658, 710

\bibitem[{{Roberts} \& {Haynes}(1994)}]{RH94}
{Roberts} M.~S., {Haynes} M.~P., 1994, \araa, 32, 115

\bibitem[{{Schweizer}(1987)}]{schweizer}
{Schweizer} F., 1987, in Nearly Normal Galaxies. From the Planck Time to the
  Present, {Faber} S.~M., ed., pp. 18--25

\bibitem[{{Spitler} {et~al.}(2008){Spitler}, {Forbes}, {Strader}, {Brodie}, \&
  {Gallagher}}]{spitler}
{Spitler} L.~R., {Forbes} D.~A., {Strader} J., {Brodie} J.~P., {Gallagher}
  J.~S., 2008, \mnras, 385, 361

\bibitem[{{van den Bergh}(1979)}]{vdb79}
{van den Bergh} S., 1979, \apj, 230, 95

\bibitem[{{Weidner} {et~al.}(2004){Weidner}, {Kroupa}, \& {Larsen}}]{weidner}
{Weidner} C., {Kroupa} P., {Larsen} S., 2004, \mnras, 350, 1503

\bibitem[{{Weilbacher} {et~al.}(2003){Weilbacher}, {Duc}, \&
  {Fritze-v.~Alvensleben}}]{weilbacher03}
{Weilbacher} P.~M., {Duc} P.-A., {Fritze-v.~Alvensleben} U., 2003, \aap, 397,
  545

\bibitem[{{Wetzstein} {et~al.}(2007){Wetzstein}, {Naab}, \&
  {Burkert}}]{wetzstein}
{Wetzstein} M., {Naab} T., {Burkert} A., 2007, \mnras, 375, 805

\bibitem[{{Whitmore} {et~al.}(1993){Whitmore}, {Schweizer}, {Leitherer},
  {Borne}, \& {Robert}}]{whitmore}
{Whitmore} B.~C., {Schweizer} F., {Leitherer} C., {Borne} K., {Robert} C.,
  1993, \aj, 106, 1354

\end{thebibliography}
\end{document}